\documentclass[a4]{elsart5p}

\usepackage{amssymb,amsmath} 
\usepackage{bm,hyphenat,xspace} %,draftcopy}
\usepackage{graphicx,epsfig}
\usepackage{tabularx}
\usepackage{epsfig} %,pstricks,psfig,epic,eepic,pst-coil}
\newcommand{\email}[1]{\ead{#1}}

\newcommand {\mbf}[1]{{\mathbf{#1}}}

\begin{document}
%\linenumbers

\begin{frontmatter}

\title{Faddeev-type calculation of nonelastic breakup in deuteron-nucleus scattering}

\author{A.~Deltuva},
\email{arnoldas.deltuva@tfai.vu.lt}

%\affiliation{
\address{
Institute of Theoretical Physics and Astronomy, 
Vilnius University, Saul\.etekio al. 3, LT-10257 Vilnius, Lithuania}
%Centro de F\'{\i}sica Nuclear da Universidade de Lisboa, 
%P-1649-003 Lisboa, Portugal }

%\received{May 31, 2022}

%\pacs{24.10.-i, 21.45.-v, 25.45.Hi, 25.40.Hs}

\begin{abstract}
The nonelastic breakup (NEB), one of channels in $(d,p)$ inclusive reactions, is studied using the
Faddeev-type  scattering theory. The NEB differential cross section 
is obtained in terms of the imaginary part of the neutron-nucleus optical potential sandwiched between the 
Alt-Grassberger-Sandhas three-body transition operators. The momentum-space calculations including
the Coulomb force are extended to higher charge numbers.
Well converged numerical results are obtained for the energy distribution of the NEB cross section,
being roughly  consistent with previous works. The spin-dependent interaction terms do not play a significant role.
The optical potential nonlocality  effect shows up at higher
proton energies, but is comparable to local potential uncertainties.
\end{abstract}


\end{frontmatter}
% \maketitle

% -----------------

\section{ Introduction} 

The collisions of two nuclei are highly complicated many-body problems,  whose complexity led to the
development of simplified approaches with rather few effective degrees of freedom. The simplest of them
is the introduction of  optical potentials that by an appropriate fit of parameters
can be quite successful in reproducing the elastic scattering observables and total reaction cross section for 
two colliding nuclei. The picture becomes more complicated if at least one of the involved nuclei
is weakly bound, such as the deuteron ($d$) or halo nucleus. The need to account for the breakup possibility  
leads to an effective three-body problem at least. For the deuteron induced reactions the active degrees of freedom
are the proton ($p$), neutron ($n$) and the involved nucleus $A$, that is most often treated as an inert particle,
whereas its compositeness is effectively accounted for via nucleon-nucleus optical potentials.
Within such a model space the elastic scattering, deuteron breakup and stripping (nucleon transfer) reactions have been
calculated using a number of approximate \cite{timofeyuk:20a,austern:87}
and rigorous three-body methods \cite{faddeev:60a,alt:67a}.
All these calculations refer to a class of processes where the nucleus $A$ remains in its ground state.
The inclusion of few lowest excited states and the respective reaction channels
have been achieved by formulation of the problem in an extended Hilbert space
\cite{summers:06a,mukhamedzhanov:12a,deltuva:13d,gomez:17a}.

However,  attempts have also been made to calculate the reactions beyond the explicitly considered model space, i.e.,
when following the deuteron breakup $d+A \to p + X$
only one of the nucleons (proton) is detected, while  another nucleon and
the nucleus $A$ can be in any state $X$, including disintegration of $A$. It is called the inclusive
$(d,p)$ breakup with  contributions from several different reaction mechanisms:
the elastic breakup (EB), nonelastic breakup (NEB), and
preequilibrium and compound nucleus (PE+CN). The latter typically is modeled using the  
statistical Hauser–Feshbach theory \cite{Wang_2011} and is beyond the scope of the present work.
The EB is the standard three-body breakup with  $A$ remaining in its ground state, while NEB includes
the breakup with simultaneous excitation of $A$ and inelastic $n+A$ processes. Both in EB and NEB three particles ($A+p+n$)
are involved explicitly, thus, the three-body calculations at least are needed. 

Under the assumption that the detected particle  acts as a spectator, i.e., predominantly scatters from $A$ elastically,
several approaches by Udagawa and Tamura \cite{PhysRevC.24.1348},
Kasano and Ichimura \cite{KASANO198281},
Hussein and McVoy \cite{HUSSEIN1985124}, and Ichimura, Austern and Vincent \cite{austern:87,PhysRevC.32.431}
have been proposed to calculate the NEB semi-inclusive differential cross section for the detected particle.
All of them explicitly include only the ground state of the nucleus $A$ and 
rely on the closure method to  sum up implicitly all final states of the subsystem $X$.
Although a formal derivation of the scattering wave functions  has been performed also
using rigorous Faddeev theory  \cite{HUSSEIN1990269}, most practical calculations, e.g.,  by
Jia et al. \cite{Wang_2011}, Moro and Lei \cite{PhysRevC.92.044616,moro-neb-fbs15}, Potel et al.
\cite{potel-neb-17}, Liu et al. \cite{PhysRevC.108.014617}, Torabi and Carlson \cite{Torabi_2023}
so far used three-body scattering wave functions based on
the distorted-wave Born-approximation (DWBA) or Glauber model. Recently also the continuum-discretized
coupled-channel method has been used \cite{lei-neb-prl,lei-neb-prc}.
In particular, Ref.~\cite{moro-neb-fbs15} pointed out that 
``the solution of the Faddeev equations is too complicated for many practical applications and, even if this solution is
available, its implementation for NEB will be very challenging.''

Experiments performed for the inclusive deuteron breakup, e.g., \cite{PhysRevC.19.370,PhysRevC.63.014610},
aimed to determine the yield of various particles ($p$, $n$, $\alpha$, ...). The collisions of deuterons with nuclei,
possible both in direct and inverse kinematics, also enable an indirect study of nucleons capture by nuclei;
see Ref.~\cite{potel-neb-17} for a recent overview. Theoretical and experimental studies of the  inclusive $(d,p)$ processes
revealed that at low proton energies the PE+CN mechanism totally dominates the differential cross section with
small contributions from EB and NEB.
This kinematic condition implies also low relative proton-nucleus energy, corresponding to long interaction time
in the classical picture, and is incompatible with the proton spectator assumption.
In contrast, at medium to highest proton energies
the PE+CN mechanism yields very small cross section whereas the NEB dominates, with a moderate contribution of EB as well.
The theoretical predictions \cite{Wang_2011,PhysRevC.92.044616,moro-neb-fbs15,potel-neb-17,PhysRevC.108.014617,Torabi_2023}
are successful in reproducing the experimental data qualitatively, while at the quantitative level some discrepancies
remain, also between different theoretical approaches for NEB. This calls for new independent methods that could possibly resolve the existing
disagreements.

Given previous successful practical applications of the Faddeev-type theory
to elastic, inelastic, transfer, and breakup reactions in three-body systems
\cite{deltuva:07d,deltuva:13d},
its extension to NEB appears
to be relevant and timely. In contrast to Ref.~\cite{HUSSEIN1990269}  that considered wave functions, in this work
the Alt, Grassberger, and Sandhas (AGS) equations  \cite{alt:67a} for three-body transition operators will be employed;
the calculations proceed in the momentum space.
This framework offers one more advantage, namely, a convenient  inclusion of nonlocal optical potentials
that have a significant impact in one-nucleon transfer reactions \cite{deltuva:09b,deltuva:23a}
and deuteron inelastic scattering \cite{deltuva:23b}. The coordinate space approaches can include the nonlocality via integro-differential
equations as well, but this has not been performed yet in the context of NEB.
Thus, after establishing the calculational scheme
for NEB, the optical potential nonlocality effect will be evaluated as well.

Another key feature of momentum-space calculations \cite{deltuva:09b,deltuva:23a,deltuva:23b} is the inclusion
of the Coulomb force via the method of screening and renormalization
\cite{taylor:74a,semon:75a,alt:80a,deltuva:05a}.
It works very well for light nuclei but becomes tedious for high charges or at very low energies,
such that practical applications are limited so far  to ${}^{58}$Ni \cite{deltuva:07d}.
An extension to higher charge numbers is desirable, especially in the view of NEB where
majority of the data refer to heavier nuclei.

Section \ref{sec.th} contains the AGS three-body scattering formalism. It shows how the extended version simulates the processes
beyond the standard three-body space and includes NEB.
The Appendix \ref{app} presents an alternative derivation of the same result using a many-body approach.
 Section \ref{sec:res} contains the results, while Sec.~\ref{concl} collects conclusions.
 
\section{Three-body transition operators and NEB \label{sec.th}}

The AGS equations \cite{alt:67a} represent  the integral equation formulation of the three-body
Faddeev scattering theory \cite{faddeev:60a} for transition operators $U_{\gamma\beta}$ that are directly related
to scattering amplitudes; Greek subscripts label the initial and final spectator particle (or the remaining pair in the
odd-man-out notation), whereas $\gamma=0$ stands for  three free particles.
In a number of previous works  AGS equations have been used for the description of
reactions with three inert particles \cite{deltuva:07d,deltuva:09b} while later extensions include
also the dynamical excitations of the core nucleus \cite{deltuva:13d,deltuva:23a,deltuva:23b}.
Notably, both versions can be cast into the same form of equations, i.e.,
\begin{equation}  \label{eq:Uba}
U_{\gamma \beta }  = (1-{\delta}_{\gamma\beta}) \, G^{-1}_{0}  +
\sum_{\alpha=1}^3   (1-{\delta}_{\alpha \gamma}) \, t_{\alpha} \, G_{0} U_{\alpha \beta}
\end{equation}
where $G_0$ is the free resolvent, and the pairwise interaction potentials  $v_{\alpha}$ enter via the respective two-particle transition operators
\begin{equation}  \label{eq:t}
t_{\alpha}  = v_{\alpha} + v_{\alpha} G_0 \, t_{\alpha}.
\end{equation}
If  excitations of the nucleus $A$ are allowed, all operators in Eq.~(\ref{eq:Uba})
become multicomponent operators acting in an extended Hilbert space
with multiple sectors \cite{deltuva:13d} corresponding to different internal states of the nucleus $A$.
The basis states 
$|\mbf{p}_\alpha \mbf{q}_\alpha a \rangle \equiv |\mbf{p}_\alpha\rangle \otimes |\mbf{q}_\alpha\rangle \otimes |a \rangle$
for the relative motion
are characterized by  Jacobi momenta for the  pair ($\mbf{p}_\alpha$)  and spectator ($\mbf{q}_\alpha$), plus the  label $a$
for the internal state of the nucleus $A$  with the excitation energy $E_a$, i.e., $E_a=0$ for the ground state.
 While previous studies with explicit core excitation included only few lowest bound states of $A$,
 formally also the continuum states corresponding to the breakup of $A$ can be discretized and included into the set $\{a\}$
 as pseudostates. 
 With a finite but sufficiently large number of such states (pseudostates)
 one may expect to account accurately for the $A$ continuum.
 %even though the practical  solution becomes impossible due to a huge number of coupled equations.
 In contrast, no discretization is used for
 the $A+p+n$ three-body continuum, described by two continuous variables,
 the Jacobi momenta $\mbf{p}_\alpha$ and $\mbf{q}_\alpha$.
 The explicit multicomponent equations can be obtained  from Eq.~(\ref{eq:Uba}) inserting the completeness relation
 \begin{equation}  \label{eq:1} 
 1 = \sum_a \int d^3\mbf{p}_\alpha d^3\mbf{q}_\alpha \,  |\mbf{p}_\alpha \mbf{q}_\alpha a \rangle
   \langle \mbf{p}_\alpha \mbf{q}_\alpha a | .
 \end{equation}
   The neutron-proton interaction and the multicomponent free resolvent
 are diagonal with respect to the different  Hilbert sectors, e.g.,
 \begin{equation}  \label{eq:g0}   
\langle a' |G_0 |a \rangle = \delta_{a'a} \, (E+i0 - H_0 - E_a)^{-1}.
 \end{equation}
 Here $E$ is the available system energy in the center-of-mass (c.m.) frame and $H_0$ the kinetic energy operator for the
 relative motion of the three clusters $A+p+n$, with eigenvalues
 $p_\alpha^2/2\mu_\alpha + q_\alpha^2/2M_\alpha$, where $\mu_\alpha$ and $M_\alpha$ are the pair and spectator reduced masses.
   In contrast,
the real nucleon-nucleus potentials  $v_{\alpha}$ couple the  different Hilbert sectors, with nonvanishing
  $\langle a' |v_{\alpha} | a\rangle$ for any combination of $a'$ and $a$. As a consequence, the  nucleon-nucleus
 transition operators in  Eq.~(\ref{eq:t}) and three-body transition operators in Eq.~(\ref{eq:Uba}) couple the
 different Hilbert sectors.  Though the above equations are of the three-body type, the many-body character of the problem
 resides in the multicomponent nucleon-nucleus potentials $\langle a' |v_{\alpha} | a\rangle$, whose microscopic calculation would
 require the solution of the many-body problem. The solution of the multicomponent three-body equations (\ref{eq:Uba})
 may be also highly challenging, if large number of  states $a$ is included. However, both difficulties can be avoided
 in the calculations of NEB cross section as will be shown in the following.

 The breakup operator is a special case of Eq.~(\ref{eq:Uba}) with $\gamma=0$, i.e.,
% amplitudes for transitions to those states are given by on-shell matrix elements of the breakup operator
 \begin{subequations}  \label{eq:u0}   
   \begin{align}
     U_{0\beta} = {} & G^{-1}_{0}  + \sum_{\alpha=1}^3   \, t_{\alpha} \, G_{0} U_{\alpha \beta} \\ \label{eq:u0b}
 = {}&  {\delta}_{\beta\alpha}  G^{-1}_{0} + (1+t_{\alpha} \, G_{0}) U_{\alpha \beta},
\end{align}
\end{subequations}
 whereas the last equation is valid for any $\alpha=1, 2,$ or 3.
 For the calculation of physical observables the operator (\ref{eq:u0}) has to be sandwiched between the
 initial  and final channel states. 
 The reaction is initiated by the collision of particle 1 and the  bound pair of particles (23)
 with energy $\epsilon_1 < 0$, 
 the initial relative momentum being $\mbf{q}_1^i$ and the available energy $E=({q}_1^i)^2/2M_1 +\epsilon_1$.
  Thus, the initial channel state is given by the direct  product of
 the bound-state wave function $|\phi_1\rangle$ and plane wave  $|\mbf{q}_1^i\rangle$, $A$ being in its ground state.
 The final breakup state could be expressed in any of the Jacobi configurations,
 but for the detected particle $\alpha$ the most convenient choice is
 $|\mbf{p}_\alpha \mbf{q}_\alpha a \rangle$.
 The corresponding semi-inclusive differential cross section  is obtained from the
 fully exclusive one \cite{deltuva:24b} via 
 summation and integration over all states of the undetected particles, i.e.,
 \begin{gather}  \label{eq:d3s}
\begin{split}
\frac{d^3\sigma_b}{d^3\mbf{q}_\alpha} = {} & {} (2\pi)^4 \, \frac{M_1}{f_s \,q_1^i} \,
\sum_{a} \int d^3\mbf{p}_\alpha \,
\delta \left(E-\frac{p_\alpha^2}{2\mu_\alpha} - \frac{q_\alpha^2}{2M_\alpha} -E_a \right) \,
\\ & \times
|\langle \mbf{p}_\alpha \mbf{q}_\alpha a | U_{01} | \phi_1 \mbf{q}_1^i \rangle|^2.
\end{split}
\end{gather}
 %with $\mu_\alpha$ and $M_\alpha$ being the associated reduced masses.
 The summation is performed also over all initial and final spin states, but for the notational brevity
 it is not explicitly indicated; instead, the
 presence of $1/f_s$ in equations will indicate the performed spin summation. The initial-state spin factor
 $f_s = (2s_1+1)(2S_1+1)$  takes care of the spin averaging,
 with $s_1$ ($S_1$) being the spin of the initial-state spectator (pair).
 Note that due to the  energy conservation $q_\alpha^2 \leq 2 M_\alpha E$.
 Applying the Eq.~(\ref{eq:u0b}) and
 the general relation $\langle \mbf{p}_\alpha a |(1+t_{\alpha} \, G_{0}) = \langle \psi^-(\mbf{p}_\alpha) a |$
 between the two-particle transition matrix and scattering wave function  %$\langle \psi^-(\mbf{p}_\alpha) a |$,
 one gets
 \begin{gather}  \label{eq:upsi}
%   \begin{split}
     \langle \mbf{p}_\alpha \mbf{q}_\alpha a | U_{01} | \phi_1 \mbf{q}_1^i \rangle
     = {}  {} \langle \psi^-(\mbf{p}_\alpha) \mbf{q}_\alpha a |U_{\alpha 1}| \phi_1 \mbf{q}_1^i \rangle .
%\end{split}
\end{gather}
 The first term in Eq.~(\ref{eq:u0b}) does not contribute assuming
 that the detected particle $\alpha$ is not free in the initial state, i.e., $\alpha \neq 1$.
 Further manipulations are similar to those of previous works
 \cite{PhysRevC.24.1348,KASANO198281,HUSSEIN1985124,austern:87,PhysRevC.32.431}, i.e.,
 the $\delta$-function in Eq.~(\ref{eq:d3s}) is rewritten as the imaginary part of the energy denominator
 according to  $\delta (x) =  -(1/\pi) \mathrm{Im} (x + i0)^{-1}$, resulting in
 \begin{gather}  \label{eq:d3s-im}
\begin{split}
\frac{d^3\sigma_b}{d^3\mbf{q}_\alpha} = {} & {}  -(2\pi)^4 \, \frac{M_1}{\pi f_s q_1^i} \, \mathrm{Im} \big[
\sum_{a} \int 
  \langle \phi_1 \mbf{q}_1^i | U_{\alpha 1}^\dagger  |\psi^-(\mbf{p}_\alpha) \mbf{q}_\alpha a \rangle \, \\ & \times
\frac{d^3\mbf{p}_\alpha}{E + i0 -\frac{p_\alpha^2}{2\mu_\alpha} - \frac{q_\alpha^2}{2M_\alpha} -E_a }
\langle \psi^-(\mbf{p}_\alpha) \mbf{q}_\alpha a |
U_{\alpha 1}| \phi_1 \mbf{q}_1^i \rangle  \big].
\end{split}
 \end{gather}

 The differential cross section for
 nucleon transfer reactions, where in the final state the undetected nucleon is captured by the nucleus $A$
 to one of the bound  states  $|\phi_\alpha^j\rangle$ with energy $\epsilon_\alpha^j$, is given as
 \cite{deltuva:23a}
 \begin{gather}  \label{eq:d3st}
\begin{split}
\frac{d^3\sigma_t}{d^3\mbf{q}_\alpha} = {} & {} (2\pi)^4 \, \frac{M_1}{f_s \,q_1^i} \,
\sum_{j} \delta \left(E - \frac{q_\alpha^2}{2M_\alpha} -\epsilon_\alpha^j \right) \,
\\ & \times
|\langle  \phi_\alpha^j \mbf{q}_\alpha  | U_{\alpha 1} | \phi_1 \mbf{q}_1^i \rangle|^2.
\end{split}
 \end{gather}
 Note, the summation is over the bound state label $j$, since $|\phi_\alpha^j\rangle$ may contain several components with different
 $a$ \cite{deltuva:23a}.
 After rewritting the $\delta$-function as in  Eq.~(\ref{eq:d3s-im}) it becomes
  \begin{gather}  \label{eq:d3st-im}
\begin{split}
\frac{d^3\sigma_t}{d^3\mbf{q}_\alpha} = {} & {}  -(2\pi)^4 \, \frac{M_1}{\pi f_s q_1^i} \, \mathrm{Im} \big[
\sum_{j} 
  \langle \phi_1 \mbf{q}_1^i | U_{\alpha 1}^\dagger  |\phi_\alpha^j \mbf{q}_\alpha \rangle \, \\ & \times
\frac{1}{E + i0  - \frac{q_\alpha^2}{2M_\alpha} -\epsilon_\alpha^j }
\langle \phi_\alpha^j \mbf{q}_\alpha |
U_{\alpha 1}| \phi_1 \mbf{q}_1^i \rangle  \big].
\end{split}
 \end{gather}

  %As in Refs.~\cite{PhysRevC.24.1348,KASANO198281,HUSSEIN1985124,austern:87,PhysRevC.32.431} one can easily
 The channel resolvent embedded into the three-body space 
 \begin{gather}  \label{eq:xga}
   \begin{split}
     G_\alpha =  {} & {} \sum_{a} \int 
    \frac{ |\psi^-(\mbf{p}_\alpha) \mbf{q}_\alpha a \rangle \,
      d^3\mbf{p}_\alpha \, \langle \psi^-(\mbf{p}_\alpha) \mbf{q}_\alpha a |}
      {E + i0 -\frac{p_\alpha^2}{2\mu_\alpha} - \frac{q_\alpha^2}{2M_\alpha} -E_a } \\
       {} & {} +  \sum_{j} 
\frac{|\phi_\alpha^j \mbf{q}_\alpha \rangle \, \langle \phi_\alpha^j \mbf{q}_\alpha |}
{E + i0  - \frac{q_\alpha^2}{2M_\alpha} -\epsilon_\alpha^j }
\end{split}
 \end{gather}
 has  continuum and bound-state  contributions that can be easily identified in Eqs.~(\ref{eq:d3s-im})
 and (\ref{eq:d3st-im}), respectively. The full cross section for the detected particle $\alpha$, being the sum
 of (\ref{eq:d3s-im})  and (\ref{eq:d3st-im}), becomes
 \begin{gather}  \label{eq:d3s-ga}
%\begin{split}
\frac{d^3\sigma}{d^3\mbf{q}_\alpha} =   -(2\pi)^4 \, \frac{M_1}{\pi f_s q_1^i} \, \mathrm{Im} \big[
%\sum_{m_s} 
  \langle \phi_1 \mbf{q}_1^i | U_{\alpha 1}^\dagger  G_\alpha U_{\alpha 1}| \phi_1 \mbf{q}_1^i \rangle_q  \big].
%\end{split}
 \end{gather}
 The subscript $q$ for the matrix element in Eq.~(\ref{eq:d3s-ga}) indicates that the intermediate variable
 $\mbf{q}_\alpha$ is not integrated out, i.e., it is a matrix element with respect to the two-body subspace
 of the undetected pair only. Kinematically the nucleon transfer and breakup reactions remain separated,
 as they correspond to $q_\alpha$ values above and below $2 M_\alpha E$, respectively.

 A simple interpretation of Eq.~(\ref{eq:d3s-ga}) is that the reaction  proceeds via
 transfer of the particle $\beta\neq\alpha$ from the bound pair (23) into the continuum of the subsystem ($1\beta$),
 with subsequent rescattering in that subsystem, whereas the particle $\alpha$ acts as a spectator.
 Thus, under the $\alpha$ spectator assumption one may expect a reasonable description of the cross section 
 also when using a simplified model space.
% As the realistic multichannel three-body calculations are far too complicated,
 In the standard three-body approach to deuteron-nucleus reactions  only the ground state of the nucleus $A$
 is included explicitly, thus, all involved operators are projected onto the ground state of $A$,
 while other states are approximately accounted for via optical potentials $v_\alpha$. The label $a$ becomes
 redundant and will be omitted, implying that states $|\mbf{p}_\alpha \mbf{q}_\alpha \rangle$
 refer to this simplified model space to be considered in the following.
 The reaction is then described by the AGS equations (\ref{eq:Uba}) with the single-component version of the
  free resolvent (\ref{eq:g0}), while the channel resolvent becomes
 \begin{subequations}     
\begin{align}  \label{eq:ga}   
  G_\alpha =  {}& (E+i0 - H_0 - v_\alpha)^{-1} \\ \label{eq:gab}   
  = {}& G_0 + G_0 t_\alpha G_0.
\end{align}
 \end{subequations}
 In particular,
 \begin{subequations}     
\begin{align}  \label{eq:imga}   
  \mathrm{Im} G_\alpha \equiv  {}& (1/2i) \{G_\alpha^\dagger[(G_\alpha^\dagger)^{-1} - G_\alpha^{-1}]G_\alpha \}
  \\ 
  = {}& -\pi (1 + G_0^\dagger t_\alpha^\dagger)  %\mathrm{Im} (G_0)
\delta(E-H_0)
  (1 + t_\alpha G_0)
   +  G_\alpha  {w_\alpha} G_\alpha, \label{eq:imgab}
\end{align}
 \end{subequations}
 where $w_\alpha = \mathrm{Im} v_\alpha$ is the imaginary part of the optical potential
 acting within the undetected pair.
 The relation (\ref{eq:imgab}) in a slightly different but equivalent form has been obtained in previous works, e.g.,
\cite{KASANO198281}, and shown to be essential in disentangling EB and NEB. In particular,
 %Given the fact that $ \mathrm{Im} G_0 = -\pi \delta(E-H_0)$,
 the first term of Eq.~(\ref{eq:imgab})  together with (\ref{eq:u0b}) leads to the cross section (\ref{eq:d3s})
 contribution
 \begin{gather}  \label{eq:d3s-eb}
\begin{split}
\frac{d^3\sigma_{\rm EB}}{d^3\mbf{q}_\alpha} = {} & {} (2\pi)^4 \, \frac{M_1}{f_s \,q_1^i} \,
%\sum_{m_s}
\int d^3\mbf{p}_\alpha \,
\delta \left(E-\frac{p_\alpha^2}{2\mu_\alpha} - \frac{q_\alpha^2}{2M_\alpha}\right) \,
\\ & \times
|\langle \mbf{p}_\alpha \mbf{q}_\alpha | U_{01} | \phi_1 \mbf{q}_1^i \rangle|^2,
\end{split}
\end{gather}
 which is the standard breakup cross section with the nucleus $A$ remaining in its ground state, i.e.,
 the elastic breakup.
 The second  term of Eq.~(\ref{eq:imgab}) together with (\ref{eq:gab}) and (\ref{eq:u0b}) leads to
 NEB cross section
\begin{subequations}  \label{eq:d3s-neb}
\begin{align}  \label{eq:d3s-neb1}
%\begin{split}
\frac{d^3\sigma_{\rm NEB}}{d^3\mbf{q}_\alpha} = {}&   -(2\pi)^4 \, \frac{M_1}{\pi f_s q_1^i} 
%\sum_{m_s} 
\langle \phi_1 \mbf{q}_1^i | U_{0 1}^\dagger G_0^\dagger w_\alpha G_0 U_{0 1}| \phi_1 \mbf{q}_1^i \rangle_q \\ \nonumber
  = {}& -(2\pi)^4 \, \frac{M_1}{\pi f_s q_1^i} 
%\sum_{m_s} 
\int [ d^3\mbf{p}'_\alpha d^3\mbf{p}_\alpha w_\alpha(\mbf{p}'_\alpha,\mbf{p}_\alpha) \\ & \times
\langle \mbf{p}'_\alpha \mbf{q}_\alpha | G_0 U_{01} | \phi_1 \mbf{q}_1^i \rangle^\ast
\langle \mbf{p}_\alpha \mbf{q}_\alpha | G_0 U_{01} | \phi_1 \mbf{q}_1^i \rangle]. \label{eq:d3s-neb2}
  %\end{split}
\end{align}
\end{subequations}

The obtained NEB differential cross section (\ref{eq:d3s-neb}), consistently with previous works
\cite{PhysRevC.24.1348,KASANO198281,HUSSEIN1985124,austern:87,PhysRevC.32.431}, is given by the
imaginary part of the optical potential for the undetected pair. The difference lies in the calculation of
scattering wave functions, that in the present work are given in terms of three-body transition operators.
While EB requires only the on-shell elements of the breakup operator  (\ref{eq:u0}),
the NEB involves integration of its half-shell matrix elements.

The AGS equations (\ref{eq:Uba}) are solved numerically in the momentum-space partial-wave representation.
The orbital angular momenta up to 3, 12 and 22 are included for the neutron-proton, neutron-nucleus, and
proton-nucleus pairs, respectively, with total angular momentum up to 35.
See Refs.~\cite{deltuva:07d,deltuva:09b,deltuva:23a,deltuva:23b,deltuva:24b} for further details.
The Coulomb interaction is included via the screening and renormalization method
\cite{taylor:74a,semon:75a,alt:80a,deltuva:05a}. In fact, the renormalization factors cancel in the
expressions for the cross section. Nevertheless, one has to ensure that the screening radius is large enough
and the results become practically independent of it; this will be demonstrated in the next section.

\section{Results \label{sec:res}}

The primary goals of the present work are to establish the calculation of NEB in the
framework of rigorous three-body AGS equations and to evaluate the optical potential
nonlocality effect.
The comparison with the experimental data would involve the preequilibrium and compound nucleus
contributions that are beyond the reach of the present calculations and therefore is left to future studies.
The most convenient physical observable for this investigation is the distribution of the energy
$E_\alpha = q_\alpha^2/2m_\alpha$
of the detected particle in the three-body center-of-mass (c.m.) frame,
\begin{gather}  \label{eq:d1s}
  \frac{d\sigma_{\rm NEB}}{dE_\alpha} = m_\alpha q_\alpha \int d^2\hat{\mbf{q}}_\alpha
  \frac{d^3\sigma_{\rm NEB}}{d^3\mbf{q}_\alpha}.
\end{gather}
This observable can be calculated directly in partial waves, performing the angular integration within
Eq.~(\ref{eq:d3s-neb1}). In the considered case of deuteron-nucleus scattering the detected
particle is assumed to be the proton, though the whole formalism could be applied equally well also to
$(d,n)$ inclusive reactions.

A realistic neutron-proton interaction with spin-orbit and tensor terms such as CD Bonn \cite{machleidt:01a}
can be included into solution of the AGS equations in the partial-wave representation. However,
it was found that the differential cross section (\ref{eq:d1s}) is insensitive to fine details
of the potential, especially of its noncentral part, therefore a simple Gaussian potential
from Ref.~\cite{austern:87} 
reproducing the deuteron binding energy and low-energy ${}^3S_1$ phase shift was adopted;
it has been used in many reaction calculations. Likewise, the optical nucleon-nucleus potentials
are taken without noncentral parts, which allows the omission of spin degrees of freedom  and leads
to a significant reduction in the number of  partial waves and calculation time. Few examples
with the full spin-dependent interaction will be shown as well.
Several parametrizations of the nucleon-nucleus optical potential are used, including those
by Koning and Delaroche (KD) \cite{koning}, Weppner et al. \cite{weppner:op},  Watson et al.~\cite{watson},
Chapel Hill (CH89) group \cite{CH89}, and 
Giannini and Ricco (GR) \cite{giannini}. The latter has nonlocal and its equivalent local versions,
to be used to estimate the nonlocality  effect. Except for the nonlocal one \cite{giannini},
the other potential contain energy-dependent  parameters. In the present study they are taken
at half the deuteron energy, a standard choice in the calculations of elastic scattering and breakup.
It is perhaps too naive for NEB where one could consider also explicit energy dependence for 
$w_\alpha$ in Eq.~(\ref{eq:d3s-neb}), but should be fully sufficient for the purpose of the present work,
i.e., establishing the calculational scheme and evaluating the nonlocality  effect.

\begin{figure}[!]
\includegraphics[scale=0.80]{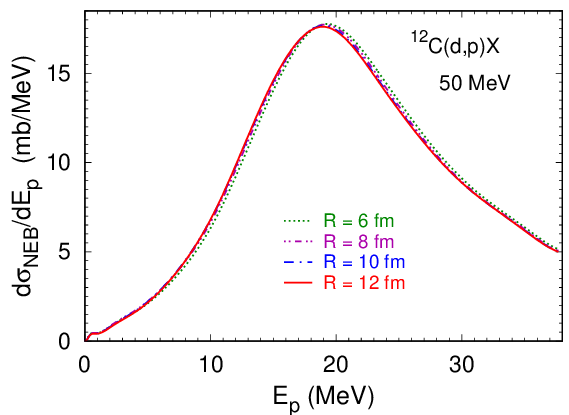} \\
\includegraphics[scale=0.80]{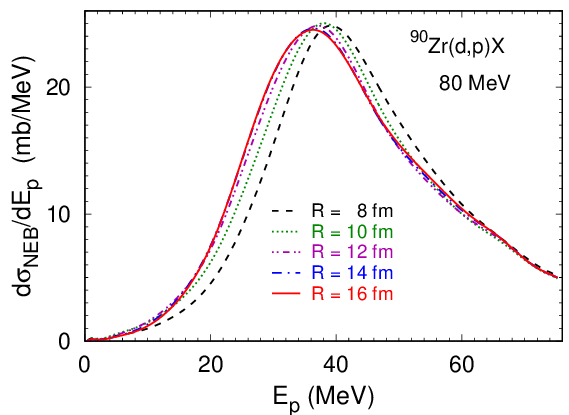} \\
\caption{\label{fig:R} (Color online)
  The convergence of the  semi-inclusive NEB cross section with the Coulomb screening radius $R$.
  50 MeV (80 MeV) deuteron scattering from ${}^{12}$C (${}^{90}$Zr) nucleus is shown in the top (bottom) panel.
  KD optical potential is used.
}
\end{figure}

First, the convergence of NEB cross section with respect to the Coulomb screening radius is demonstrated
in Fig.~\ref{fig:R}. The screening function $e^{-(r/R)^n}$ is taken over from Ref.~\cite{deltuva:05a}
with $n=8$, though the results are quite insensitive to $n$ between 4 and 10.
Two cases are considered, deuteron scattering from ${}^{12}$C  and ${}^{90}$Zr nuclei at
50 and 80 MeV beam energy, respectively. While for $Z=6$ target the convergence is very fast, even
$R=6$ fm is quite sufficient, the convergence becomes considerably slower for $Z=40$, where
$R$ around 15 fm at least is needed. Nevertheless, achieving well converged results for ${}^{90}$Zr
with $Z=40$ is an important step forward compared to previous Faddeev-type calculations
with the screening and renormalization approach,  limited to ${}^{58}$Ni with  $Z=28$.

%The numerical results are openly available \cite{midas}.

%%%%%%%%%%%%%%%%%%%%%%%%%%%%%%%%%%%%%%%%%%%%%%%%%%%%%%%%%%%% 

\begin{figure}[!]
\includegraphics[scale=0.80]{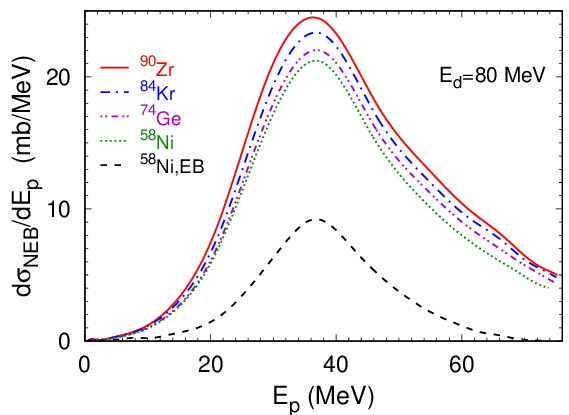}
\caption{\label{fig:nizr} (Color online)
  The semi-inclusive NEB cross section for 80 MeV deuteron scattering from
  ${}^{58}$Ni, ${}^{74}$Ge, ${}^{84}$Kr, and ${}^{90}$Zr nuclei
  as a function of the proton energy in c.m. frame.
  KD optical potential is used.
  In the case of ${}^{58}$Ni also the EB cross section is displayed by dashed curve.
}
\end{figure}

Results for 80 MeV deuteron scattering from most abundant isotopes of
$Z=28$, 32, 36, and 40 nuclei, i.e.,  ${}^{58}$Ni, ${}^{74}$Ge, ${}^{84}$Kr, and ${}^{90}$Zr
are presented in Fig.~\ref{fig:nizr}. The differential NEB cross section has  an asymmetric bell shape and
shows increase with $Z$, which is not strictly linear, probably
indicating the dependence on the neutron-proton asymmetry $(A-2Z)/A$ as well.
In the case of ${}^{58}$Ni also the corresponding EB cross section is shown, which is considerably lower
than the NEB one. The EB peak around half of the deuteron beam energy corresponds to the quasifree neutron scattering
off the nucleus $A$. As the half-shell breakup operator appears 
also in Eq.~(\ref{eq:d3s-neb}), the NEB cross section peaks in the same energy region.
These ${}^{58}$Ni results, transformed to the lab frame,  allow a comparison with previous works, e.g.,
\cite{PhysRevC.92.044616,Torabi_2023}. Their energy distributions also have an asymmetric bell shape (with few wiggles
in Ref.~\cite{PhysRevC.92.044616}),  but the Faddeev-type approach
predicts larger cross section: at the  peak the results of Refs.~\cite{PhysRevC.92.044616} and \cite{Torabi_2023} are lower by
10 and 35\%, respectively.
At the same time, as shown in Fig. 3 of Ref.~\cite{PhysRevC.92.044616},
the PE+CN contribution is sizable and even becomes dominant at lower proton energies.

\begin{figure}[!]
\includegraphics[scale=0.80]{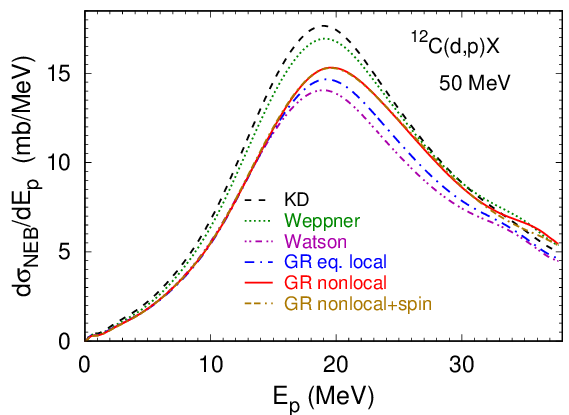} \\
\includegraphics[scale=0.80]{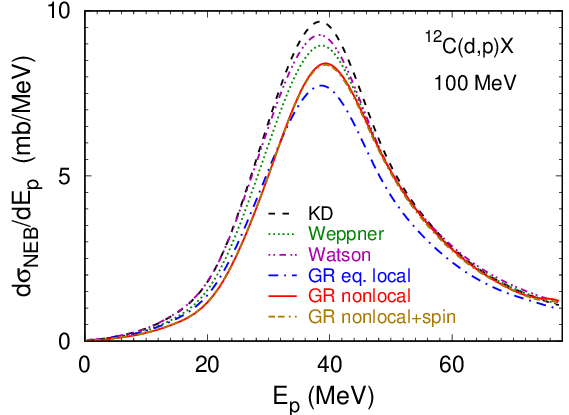} \\
\caption{\label{fig:c} (Color online)
  The semi-inclusive NEB cross section for deuteron-${}^{12}$C scattering at 50 and 100 MeV beam energy
  as a function of the proton energy in c.m. frame.
  Results with several neutron-proton and nucleon-nucleus optical potential parametrizations are compared.  
}
\end{figure}

\begin{figure}[t]
\includegraphics[scale=0.80]{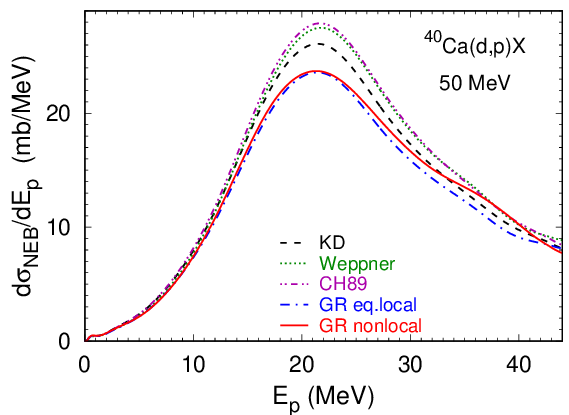} \\
\includegraphics[scale=0.80]{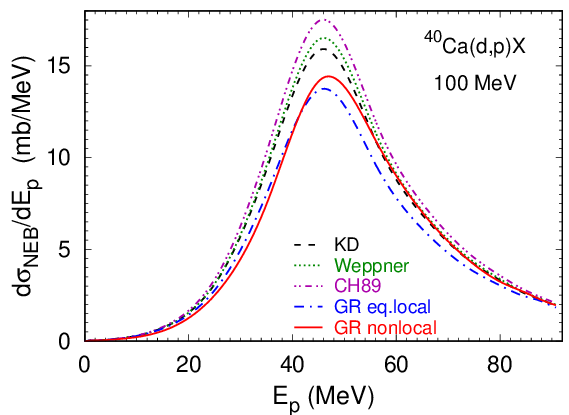} \\
\caption{\label{fig:ca} (Color online)
  The semi-inclusive NEB cross section for deuteron-${}^{40}$Ca scattering at 50 and 100 MeV beam energy
  as a function of the proton energy in c.m. frame.
  Results with several optical potential parametrizations are compared.  
}
\end{figure}

The optical potential nonlocality effect is investigated in Figs.~\ref{fig:c} and \ref{fig:ca}.
The GR potentials \cite{giannini} were developed for symmetric $N=Z$ nuclei,
and further improved for ${}^{12}$C, ${}^{16}$O and ${}^{40}$Ca \cite{deltuva:09b}
by an  additional readjustment of parameters to fit the experimental data.
The present study explores deuteron scattering from ${}^{12}$C and ${}^{40}$Ca
nuclei at 50 and 100 MeV beam energy.
In addition to the GR nonlocal optical potential and its roughly equivalent local potential, results
were obtained also with other commonly used local potentials, i.e.,
KD \cite{koning}, Weppner \cite{weppner:op},  Watson \cite{watson}, and
CH89 \cite{CH89}. The nonlocality effect, taken as the difference between local and nonlocal GR potentials,
 visibly increases the NEB cross section at the peak and beyond. On the other hand,
the sensitivity to the model of the local potential appears even more sizable.
%KD - not design, Wp - overest n12C
Noteworthy, at higher $E_p$ values the 
nonlocal GR potential results agree quite well with those of most local potentials except the GR one.
Furthermore, the shape of the  NEB cross section depends on the beam energy as well, the high-energy tail becoming lower
with increasing beam energy.

Figure \ref{fig:c} also studies the importance of spin-orbit force in the optical potential
and the realistic neutron-proton potential CD Bonn \cite{machleidt:01a}. Both
are included in the results labeled ``GR nonlocal+spin''. Quite surprisingly, the effect for NEB
cross section turns out to be almost negligible. Thus, even a very simple and unrealistic $np$ potential
might be sufficient for NEB calculations.

%The very similar shape of the differential NEB cross sections in Fig.~\ref{fig:nizr} peaking at half the maximal energy
%might suggest there is a universal curve for all nuclei.
%However, Figs.~\ref{fig:c} and \ref{fig:ca} show the dependence both on the optical potential and beam energy, most clearly seen
%at the peak and at maximal proton energy, respectively.

%\clearpage

\section{Conclusions \label{concl}}

The nonelastic breakup, comprising a class of channels in $(d,p)$ inclusive reactions, was studied using the
Faddeev-type three-body scattering theory. The key steps in deriving the NEB cross section, i.e.,
the spectator assumption for the proton, the closure over final states and the subsequent projection onto the
ground state of the nucleus $A$, are the same as in previous works. Consistently, the NEB cross section
is driven by the imaginary part of the  optical potential for the undetected pair. The difference
lies in the description of scattering states which in the present study is given in terms of the
AGS transition operators, calculated in the momentum-space partial-wave representation.

Well converged numerical results were obtained for the energy distribution of the NEB cross section.
They have asymmetric bell shape as found in  previous works  but are 10 to 35\% higher.
Benchmark calculations are left for future studies.
They are very important given the existing differences between various NEB approaches
\cite{Wang_2011,potel-neb-17,PhysRevC.108.014617,Torabi_2023}.

The effect of the optical potential nonlocality was also investigated. It is possibly more pronounced at higher
proton energies, but remains within uncertainties due to different local optical potential parametrizations.
The NEB cross section appears to be insensitive to the spin-dependent interaction terms.

Finally, performing calculations with nuclei of higher charge number, presently up to $Z=40$,
is an important step towards application of the Faddeev-type momentum-space equations
also to other reactions involving medium mass and heavy nuclei, e.g.,
the deuteron stripping $(d,p)$.
Preliminary studies \cite{djpriv} indicate that a combination with machine learning methods could be
a powerful tool for a further extension, whereas the results of direct calculations like here could serve
as the training data.

%%%%%%%%%%%%%%%%%%%%%%%%%%%%%%%%%%%%%%%%%%%%%%%%%%%%%%%%%%%%
\vspace{1mm}
%\begin{acknowledgments}
  Author thanks D. Jur\v{c}iukonis and A. M. Moro for discussions.
This work has received funding from the 
Research Council of Lithuania (LMTLT) under Contract No. S-CERN-24-2.
%\end{acknowledgments}

\begin{appendix}
  \section{Many-body approach to NEB \label{app}}
  I consider the collision of the bound pair $(\alpha\beta)$ with a bound $A$-body system in the c.m. frame,
  having the $A(d,p)X$ reaction in mind.
  This is an ($A+2$)-body problem, the Hamiltonian $H_0 + V$ contains $(A+1)$ kinetic energy contributions $H_0$
  from all relative motions 
  plus the sum of all $(A+2)(A+1)/2$   pairwise potentials, eventually supplemented by three-body and  higher forces,
  all being real.
  Let $E$ be the available   energy, $|\Psi^+(\mbf{q}_1^i)\rangle$ the full many-body scattering wave function with the
  initial relative deuteron-nucleus momentum $\mbf{q}_1^i$, and $\mbf{q}_\alpha$ the final relative momentum between
  the detected particle $\alpha$ and the remaining subsystem $X$ which can be in any state.
  
  If $X$ forms a bound nucleus  with wave function $|\phi_X^j\rangle$ and energy $\epsilon_X^j$, where $j$ labels the different
  bound states, the reaction amplitude can be given
  as $\langle \phi_X^j \mbf{q}_\alpha | V_{\alpha X} |\Psi^+(\mbf{q}_1^i)\rangle$. Here  $V_{\alpha X}$ is the interaction that is
  external to the final $\alpha+X$ channel, i.e.,  the sum of all potentials
  (two-body, three-body, etc.) acting between the particle $\alpha$ and the subsystem $X$.
  The above amplitude also defines the transfer component of the transition operator
  \begin{equation} \label{eq:uax}
    U_{\alpha X, d A} | \phi_d \phi_A \mbf{q}_1^i \rangle \equiv  V_{\alpha X} |\Psi^+(\mbf{q}_1^i)\rangle,
  \end{equation} 
  acting on the initial channel state, given as product of bound-state wave functions of deuteron and nucleus $A$
  and plane wave for their relative motion. 
  
If $X$ is broken into $n$ clusters $X_j$ with internal energies $\epsilon_j$,  $j=1,...,n$,
the reaction amplitude can be given as
$\langle \psi^-(\{X_j\},\{\mbf{p}_k\}) \, \mbf{q}_\alpha | V_{\alpha X} |\Psi^+(\mbf{q}_1^i)\rangle$.
Here $\langle \psi^-(\{X_j\},\{\mbf{p}_k\})|$ is the full wave function of the subsystem $X$ corresponding
to the set of outgoing clusters $\{X_j\}$  with reduced masses $\mu_k$ and
asymptotic relative momenta $\mbf{p}_k$,  $k=1,...,n-1$. Note that  $\mu_k$ and $\mbf{p}_k$ depend
on the clustering $\{X_j\}$, but this dependence is suppressed for brevity.
The energy of such a state is $E(\{X_j\},\{\mbf{p}_k\}) = \sum_k p_{k}^2/2\mu_k + \sum_j \epsilon_{j}$.
The differential cross section for the detected particle $\alpha$ is the sum over all possible channels,
 \begin{gather}  \label{eq:Xd3s}
\begin{split}
\frac{d^3\sigma}{d^3\mbf{q}_\alpha} = {} & {} (2\pi)^4 \, \frac{M_1}{f_s \,q_1^i} \,
\Big [ \sum_{j} 
  \delta \Big(E- \frac{q_\alpha^2}{2M_\alpha} -\epsilon_X^j \Big) \,
  \\ & \times
|\langle \phi_X^j \mbf{q}_\alpha | V_{\alpha X} |\Psi^+(\mbf{q}_1^i)\rangle|^2
\\ & +
\sum_{\{X_j\}} \int (\prod_k d^3\mbf{p}_k) \delta \Big(E- \frac{q_\alpha^2}{2M_\alpha} - E(\{X_j\},\{\mbf{p}_k\}) \Big)
\\ & \times
|\langle \psi^-(\{X_j\},\{\mbf{p}_k\}) \, \mbf{q}_\alpha | V_{\alpha X} |\Psi^+(\mbf{q}_1^i)\rangle|^2 \Big],
\end{split}
 \end{gather}
 i.e., the second sum runs over all possible clusterings.

 Rewriting the $\delta$-functions as  in  Sec.~\ref{sec.th}, one can isolate in Eq.~(\ref{eq:Xd3s})
 the resolvent for the subsystem $X$, i.e., 
 \begin{gather}  \label{eq:GX}
   \begin{split}
%     G_X = {} & {}  \sum_{j} 
 &    G_X = {}  {}  \sum_{j} 
\frac{|\phi_X^j \mbf{q}_\alpha \rangle \, \langle \phi_X^j \mbf{q}_\alpha |}
     {E + i0  - \frac{q_\alpha^2}{2M_\alpha} -\epsilon_X^j } \\
     {} & {}+ \sum_{\{X_j\}} \int (\prod_k d^3\mbf{p}_k)
     \frac{ |\psi^-(\{X_j\},\{\mbf{p}_k\}) \, \mbf{q}_\alpha \rangle  \langle \psi^-(\{X_j\},\{\mbf{p}_k\}) \, \mbf{q}_\alpha |}
{E+i0 - \frac{q_\alpha^2}{2M_\alpha} - E(\{X_j\},\{\mbf{p}_k\})}.
\end{split}
 \end{gather}

 Together with Eq.~(\ref{eq:uax}) the differential cross section (\ref{eq:Xd3s}) becomes
\begin{gather}  \label{eq:d3s-gx}
%\begin{split}
\frac{d^3\sigma}{d^3\mbf{q}_\alpha} =   -(2\pi)^4 \, \frac{M_1}{\pi f_s q_1^i} \, \mathrm{Im} \big[
%\sum_{m_s} 
  \langle \phi_d \phi_A \mbf{q}_1^i | U_{\alpha X, dA}^\dagger  G_X U_{\alpha X, dA}| \phi_d \phi_A \mbf{q}_1^i \rangle_q  \big].
%\end{split}
 \end{gather}
where, as in Sec.~\ref{sec.th},  the  subscript $q$  indicates that the matrix element in Eq.~(\ref{eq:d3s-gx}) is
taken with respect to $X$ variables only.

Finally, the three-body reduction, i.e., the projection onto the ground state of the nucleus $A$ implies the replacement
$G_X \to G_\alpha$ and $U_{\alpha X, dA}| \phi_d \phi_A \mbf{q}_1^i \rangle \to U_{\alpha 1}| \phi_1 \mbf{q}_1^i \rangle$,
leading exactly to the three-body result as in Eq.~(\ref{eq:d3s-ga}).

The above derivation is quite similar to those given in Refs.~\cite{austern:87,PhysRevC.32.431,HUSSEIN1990269},
  but does not assume the infinitely heavy nucleus $A$, and instead of the distorted proton wave introduces the transition
  operator.
  
\end{appendix}

%%%%%%%%%%%%%%%%%%%%%%%%%%%%%%%%%%%%%%%%%%%%%%%%%%%%%%%%%%%%
%\bibliographystyle{plbsty}
%\bibliography{abbrev,pre80,80-89,90-99,200x,clmb,ad,exp,book,adexp,nreact,nstruct,neb}

 \end{document}